\documentclass[12pt]{article}

\usepackage{epsfig}
\usepackage{amsmath}
\usepackage{soul,color}
\usepackage{bm}
\usepackage{epsfig}
\usepackage{natbib}

\def\b#1{\mbox{\boldmath $#1$}}    
\def\bl#1{\mbox{\footnotesize \boldmath {$#1$}}} 

\renewcommand{\th}{\theta}

\usepackage{bm}
\usepackage{epsfig}
\def\bPi{\mbox{\boldmath$\Pi$}}
\def\bPhi{\mbox{\boldmath$\Phi$}}
\begin{document}

\title{\vspace*{-1.5cm}
A comparison of some criteria for states selection in the latent Markov model for longitudinal data}
\author{Silvia Bacci\footnote{Department of Economics, Finance and Statistics,
University of Perugia, Italy; {\em email}: \;\;\;\;\;\;\;\;\;\;\;\;\;\;\;\;\;\;\;\;\;\; silvia.bacci@stat.unipg.it.},
Silvia Pandolfi$^*$\footnote{{\em email}: pandolfi@stat.unipg.it} ,
Fulvia Pennoni\footnote{Department of Statistics and Quantitive Methods, University of Milano-Bicocca,  Milan, Italy; {\em email}: fulvia.pennoni@unimib.it} } 
 \maketitle

\begin{abstract}
We compare different selection criteria to choose the number of
latent states of a multivariate latent Markov model for longitudinal
data. This model is based on an underlying Markov chain to represent
the evolution of a latent characteristic of a group of individuals
over time. Then, the response variables observed at the different
occasions are assumed to be conditionally independent given this
chain. Maximum likelihood of the model is carried out through an
Expectation-Maximization algorithm based on forward-backward
recursions which are well known in the hidden Markov literature for
time series. The selection criteria we consider in our comparison
are based on penalized versions of the maximum log-likelihood or on
the posterior probabilities of belonging to each latent state, that
is the conditional probability of the latent state given the
observed data. A Monte Carlo simulation study shows that the indices
referred to the log-likelihood based information criteria perform in
general better with respect to those referred to the classification
based criteria. This is due to the fact that the latter tend to
underestimate the true number of latent states, especially  in the
univariate case. \vspace{3mm}
\end{abstract}

\section{Introduction}
\label{intro}

A crucial element in the literature about the wide class of mixture
models \citep{mcla:peel:00} is represented by the choice of the
number of mixture components, which represents a specific aspect
related with the model selection process. For instance, this issue
arises in the context of latent class (LC) models about the choice
of latent classes and in the contexts of hidden Markov (HM) models
for time-series and stochastic processes \citep{zucc:macd:09} and of
latent Markov (LM) models \citep{Wig:73} for longitudinal data. This
last class of models is typically used when the interest is in
describing the evolution of a latent characteristic of a group of
individuals over time. They assume that one or more
occasion-specific response variables depend only on a discrete
latent variable, characterized by a given number of latent states,
which follows a first-order Markov process
\citep{bart:farc:penn:13}. The basic idea behind this assumption is
that the latent process fully explains the observable behavior of a
subject. Furthermore, the latent state to which a subject belongs to
at a certain occasion only depends on the latent state at the
previous occasion. An LM model may also be seen as an extension of
the LC model, in which the assumption that each subject belongs to
the same latent class throughout the period of observation is
suitable relaxed.

In such a context the number of latent states is usually selected on
the basis of the observed data, both in the case of the basic LM
model or in the advanced versions that, for example, allow for the
inclusion of observable individual covariates. Only in certain
applications the number of latent states is a priori defined by the
nature of the problem or by the interest of the research. However, states
selection on the basis of the observed data implies that
increasing the number of states often improves the fit of the model,
as judged by the likelihood, but also the number of parameters. The
same problem arises when selecting the number of components in a
finite mixture model.

The more common approaches which have been adopted to balance model
fit and parsimony are based on information criteria constructed
according to indices that are penalized versions of the maximum
log-likelihood. Among these criteria, the most common are the Akaike
Information Criterion \citep[AIC;][]{aka:73} and the Bayesian
Information Criterion \citep[BIC;][]{schw:78}. The first one is
known as an estimator of the Kullback-Leibler discrepancy between
the model generating the data and the fitted model. BIC may be
instead seen as an asymptotic approximation of the integrated
likelihood, which provides an estimator of a transformation of the
Bayesian posterior probability of a candidate model. Several
alternative to the AIC criterion have been proposed in literature
such as AIC$_3$ of \cite{bozdogan1993} and the Consistent AIC (CAIC)
criterion proposed by \cite{Bozdogan1987} which are based on
different penalization terms. It is important to mention that the
information criteria are preferred to methods based on the
likelihood ratio test between nested models because the latter
require bootstrap resampling procedure.

In addition to the above log-likelihood based information criteria,
classification based criteria have been proposed in literature,
which allow us to measure the quality of the classification provided
by a model. The Normalized Entropy Criterion (NEC) is an approach
first developed by \cite{cel:sor:96} to select the number of
components in the context of mixture models. It is based on an
entropy term computed on the basis of the posterior probabilities
for every sample unit and mixture component. This criterion takes
into account the quality of the classification, and then how well
the clusters are separated, further to the goodness-of-fit of the
model, which is measured in terms of log-likelihood. An entropy
index has been recently proposed in HM literature to measure
uncertainty involved in connection with finding the most likely
sequence of the latent states; see \cite{hern:cres:cybe:05} and
\cite{dura:gu:12}. The entropy measure, however, has not been
investigated as a tool for states selection in such a context. Among
other classification based criteria, it is worth mentioning also the
Classification Likelihood information Criterion (CLC), adopted by
\cite{bier:gova:97} in the mixture context, and an approximation of
the Integrated Classification Likelihood criterion
\citep[ICL;][]{bier:cele:gova:98} using BIC denoted as (ICL-BIC)
firstly adopted by \cite{bier:cele:gova:00} and \cite{mcla:peel:00}.

In the context of finite mixture models and, in particular, of LC 
models, several studies exist aimed at comparing the performance of
the above mentioned criteria. Among others, \cite{fraley:raft:02}
used BIC for clustering in mixture models, showing its satisfactory
behavior \citep[see also][Ch. 8]{mcla:peel:00}. Simulation studies
have also been performed by \cite{nyl:et:al:07} for growth mixture
and LC models, and by \cite{bier:gova:99} for Gaussian mixtures. In
both situations it was found that BIC outperforms the other
information criteria. We also refer to \cite{dias:06} for a study
refer to the LC model with binary response variables in which
emerges that AIC$_3$ is the best criterion for selecting the number
of latent classes. Moreover, CAIC has been proved to have a similar
performance with respect to BIC \citep{lin:day:97}. About the
behavior of the classification based criteria, we refer to
\cite{bier:gova:99}, which found that NEC gives poor results in
selecting the model under comparison, although it exhibits good
behavior in detecting the number of clusters. Moreover,
\cite{bier:cele:gova:00} showed that ICL appears to be more robust
than BIC to violation of some of the mixture model assumptions.

Even if these criteria are widely used in literature, their
performance have not been studied enough in detail in connection
with LM models. A comparison of AIC and BIC performance in
connection with states selection of a univariate LM model may be
found in \cite{bart:farc:penn:13}[Ch. 7]. However, to our knowledge,
there are no studies aimed at comparing the behavior of the
different information criteria mentioned above. On the other hand
their properties have been studied in the context of HM models; see,
among others, \cite{cele:dura:08}, \cite{costa:deange:10} and the
references therein. However, the context is quite different, since
HM models are used for time series, whereas LM models are applied to
longitudinal data. The main purpose of this paper is to compare the
performance of all the illustrated information criteria when applied
to select the number of latent states in a multivariate LM model.

We show a Monte Carlo simulation study on the basis of different
model specifications, with respect to the number of response
variables, and to different conditional response probabilities and
transition probabilities. The aim is to analyze the effect of these
factors on selecting the number of states and to set up a comparison
between log-likelihood based and classification based criteria. In
particular, in applying the NEC criterion, we consider an entropy
measure based on the posterior probabilities of all the possible
configurations of latent states, given the observed data, for every
sample unit. We also consider two approximations of NEC which are
based on a modified version of the entropy computed on the basis of
the posterior probability of every single latent state at every time
occasion.

The article is organized as follows. In the following section we
illustrate the multivariate LM model and we deal with maximum
likelihood estimation of the model parameters. In Section
\ref{Sec:criteria} we illustrate the  latent states selection criteria under
comparison. In Section \ref{Sec:sim} we show the results of a series
of simulations made in order to assess the quality of the analyzed
criteria. In Section \ref{sec:concl} we provide main conclusions.

\section{The multivariate LM model}\label{sec:2}
%
In the multivariate formulation of the LM model (that includes the
univariate one as a special case) we observe a vector of categorical
response variables $\b Y^{(t)} = (Y_1^{(t)}, \ldots, Y_r^{(t)})$,
for $t= 1, \ldots, T$. Each variable $Y_j^{(t)}$, $j = 1, \ldots,
r$, has $c_j$ categories, labeled from 0 to $c_j-1$. We denote by
$\b Y = (\b Y^{(1)}, \ldots, \b Y^{(T)})$ the vector of observed
responses made of the union of vectors $\b Y^{(t)}$, which usually, 
is referred to repeated measurements of the same variables $Y_j$ ($j
= 1, \ldots, r$) on the same individuals at different time points.

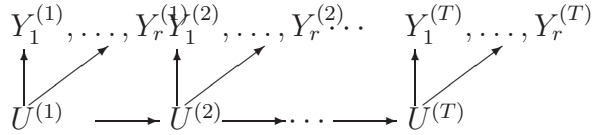
\begin{figure}[ht]\centering
\begin{picture}(150,45)
%
%
\put(-20,35){$Y_1^{(1)}, \ldots, Y_r^{(1)}$}
\put(39,35){$Y_1^{(2)}, \ldots, Y_r^{(2)}$}
\put(-20,0){$U^{(1)}$}
\put(-13,9){\vector(4,3){30}}
\put(40,0){$U^{(2)}$}
\put(-15,9){\vector(0,1){21}}
\put(12,3){\vector(1,0){24}}
\put(43,9){\vector(0,1){21}}
\put(46,9){\vector(4,3){30}}
\put(60,3){\vector(1,0){24}}
\put(85,0){$\cdots$}
\put(100,35){$\cdots$}
\put(102,3){\vector(1,0){24}}
\put(129,35){$Y_1^{(T)}, \ldots, Y_r^{(T)}$}
\put(130,0){$U^{(T)}$}
\put(133,9){\vector(0,1){21}}
\put(136,9){\vector(4,3){30}}
\end{picture}\caption{\em Path diagram of the basic latent Markov model for
multivariate data}\label{fig:path_multi}
\end{figure}

The model is based on two main assumptions. Firstly, the vectors $\b Y^{(t)}$ are
conditionally independent given a latent process $\b U = (U^{(1)},
\ldots, U^{(T)})$, and the response variables in each of these
vectors are conditionally independent given $U^{(t)}$ at time $t$
with state space $\{1, \ldots, k\}$. In other words, each occasion-specific observed
variable $Y_j^{(t)}$ is independent of $Y_j^{(t-1)}, \ldots,
Y_j^{(1)}$ and of  each $Y_h^{(t)}$, for all $h\neq j = 1, \ldots,
r$, given $U^{(t)}$.  This is the so called
\textit{local independence} assumption. Secondly,  the latent process $\b U$ is assumed to follow
a first-order Markov chain with $k$ latent states, that is each latent
variable $U^{(t)}$ is independent of $U^{(t-2)}, \ldots, U^{(1)}$,
given $U^{(t-1)}$. The resulting model is represented by the path
diagram in Figure~\ref{fig:path_multi}.


\vspace*{0.2cm}

The model is characterized by three different types of parameters:
\begin{itemize}
\item the conditional response probabilities
\[
\phi_{jy|u}^{(t)} = p(Y_j^{(t)} = y | U^{(t)} = u),
\]
with $j= 1, \ldots, r$,  $t = 1, \ldots, T$, $u= 1, \ldots, k$, and
$y = 0, \ldots, c_j-1$, which may be collected into the vector
\[
\phi_{\bl y|u}^{(t)} = \prod_{j=1}^r \phi_{jy|u}^{(t)}  =
p(Y_1^{(t)} = y_1, \ldots, Y_r^{(t)} = y_r| U^{(t)} = u);
\]
\item the initial probabilities
\[
\pi_u = p(U^{(1)} = u),
\]
with $u = 1,\ldots, k$;
\item the transition probabilities
\[
\pi_{u|v}^{(t|t-1)} = p(U^{(t)} = u | U^{(t-1)} = v),
\]
with $t = 2, \ldots, T$, $u, v = 1, \ldots, k$.
\end{itemize}
Note that all these probabilities do not depend on the specific
sample unit. Moreover, it is possible to include a constraint on the
transition probabilities corresponding to the hypothesis that the
Markov chain is time homogeneous. Under this hypothesis, which is
considered in the simulation study illustrated in
Section~\ref{Sec:sim}, the transition probabilities do not depend on
$t$, so as
\[
\pi_{u|v}^{(t|t-1)} = \pi_{u|v},\quad t = 2,\ldots,T.
\]
The number of free parameters of the multivariate LM model above is
given by
\begin{equation}\label{eq:par} \#\mathrm{par}=\underbrace{k
\sum_{j=1}^r
(c_j-1)}_{\phi_{jy|u}^{(t)}}+\underbrace{k-1}_{\pi_u}+\underbrace{(T-1)k(k-1)}_{\pi_{u|v}^{(t|t-1)}}.
\end{equation}

The probability mass function of the distribution of $\b U$ may be
expressed as
\[
p(\b U = \b u)  = \pi_{u} \prod_{t=2}^T \pi_{u| v}^{(t|t-1)}
\]
and the conditional distribution of $\b Y$ given $\b U$ is
\[
p(\b Y = \b y | \b U = \b u) = \prod_{t=1}^T \phi_{\bl y|u}^{(t)} =
\phi_{\bl y|u}^{(1)} \cdot \phi_{\bl y|u}^{(2)} \ldots \phi_{\bl
y|u}^{(T)}.
\]
Therefore, the manifest distribution $p(\b Y = \b y)$ of  $\b Y$ follows
\[
p(\b Y = \b y) =\sum_{\bl
u}\pi_{u}\pi_{u|v}^{(2|1)}\cdots\pi_{u|v}^{(T|T-1)}
\phi_{jy|u}^{(1)}\cdots \phi_{jy|u}^{(T)}.
\]
Note that computing $p(\b Y = \b y)$ involves all the possible $k^T$
configurations of vector $\b u$, that typically requires a
considerable computational effort. In order to efficiently compute
this probability we can use a forward recursion
\citep{baum:70,welch:2003} for obtaining
\[
q^{(t)}_{u, \bl y} = p(U^{(t)}=u, \b Y^{(1)}, \ldots, \b Y^{(t)}).
\]
In particular, the $t$-th iteration of this recursion, for
$t=2,\ldots,T$ consists of computing
\begin{align*}
q^{(t)}_{u, \bl y} = \sum_{v=1}^k q^{(t-1)}_{v,\bl y}
\pi_{u|v}^{(t|t-1)}  \phi_{\bl y|u}^{(t)}  \quad u =1, \ldots, k
\end{align*}
starting with $q^{(1)}_{u, \bl y} = \pi_{u} \phi_{\bl y|u}^{(1)}$,
for $t=1$. This recursion may be easily implemented by using matrix
notation; see \cite{bart:farc:penn:13} for details.

\subsection{Likelihood  inference}\label{sec:lik}
In an observed sample of $n$ subjects, let $n_{(\bl y)}$ be the
frequency of the observed response configuration  $\b y$, and
assuming independence between the sample units, the model
log-likelihood may be computed as
\[
\ell(\b\th) = \sum_{\b y} n_{(\bl y)}\log[p(\b Y = \b y)],
\]
where $\b\th$ is the vector of all model parameters arranged in a
suitable way. The model log-likelihood may be maximized with respect
to $\b\th$ by using the Expectation-Maximization (EM) algorithm of
\cite{demp:et:al:77} which represents the main tool to estimate this
class of models. This algorithm is based on the concept of {\em
complete data}, which is represented by the pair $(\b u, \b y)$,
where $\b u$ denotes a realization of $\b U$. Therefore, the
complete data log-likelihood is given by
\begin{align*}
\ell^*(\b \theta)  & = \sum_{j=1}^r\sum_{t=1}^T\sum_{u=1}^k\sum_{y=0}^{c-1} a_{juy}^{(t)} \log \phi_{jy|u}^{(t)} + \\
    &   + \sum_{u=1}^k b_u^{(1)} \log \pi_u + \sum_{t=2}^T\sum_{v=1}^k\sum_{u=1}^k b_{vu}^{(t)} \log \pi_{u|v}^{(t|t-1)}
\end{align*}
where $a_{juy}^{(t)}$ corresponds to the frequency of subjects
responding by $y$ for the $j$-th response variable and belonging to
latent state $u$ at time $t$, $b_u^{(1)}$ is the frequency of
subjects in latent state $u$ at time $1$, and $b_{vu}^{(t)}$
corresponds to the frequency of subjects which move from latent
state $v$ to state $u$ at time $t$.

Since the latent configuration for each subject is not known the EM
maximizes the log-likelihood above by alternating the following two
steps until convergence:
\begin{itemize}
\item{\bf E-step}: compute the expected value of
the above frequencies, given the observed data and the current value
of the parameters, so as to obtain the expected value of
${\ell^*}(\b\th)$
\item{\bf M-step}: update $\b\th$ by maximizing
the expected value of $\ell^*(\b \theta)$ obtained above; explicit
solutions for $\b\th$ estimation are available at this aim, see
\cite{bart:farc:penn:13}.
\end{itemize}

The  E-step of the algorithm involves the computation of the
posterior probabilities $f_{u|\bl y}^{(t)}$ and $f_{u|v, \bl
y}^{(t|t-1)}$. Using the following backward  recursion
\begin{align*}
\bar{q}^{(t)}_{v, \bl y} = p(\b Y^{(t+1)}, \ldots, \b
Y^{(T)}|U^{(t)}=v) = \sum_{u=1}^k \bar{q}^{(t+1)}_{u,\bl y}
\pi_{u|v}^{(t+1|t)}  \phi_{\bl y|u}^{(t+1)}  \quad v =1, \ldots, k,
\end{align*}
starting with  $\bar{q}^{(T)}_{v,\bl y}= 1$, for $t=1,\ldots,T$, we
have
\begin{align}
f^{(t)}_{u|\bl y}=\frac{q^{(t)}_{v, \bl y}\bar{q}^{(t)}_{u,\bl
y}}{P(\b Y = \b y)},\quad u=1,\ldots,k,
\label{eq:1}
\end{align}
whereas, for $t=2,\ldots,T$ and $u,v=1,\ldots,k$, we have 
\begin{align}
f_{u|v, \bl y}^{(t|t-1)} = \frac{q^{(t-1)}_{v,\bl
y}\pi^{(t)}_{u|v}\phi_{\bl y|u}^{(t)}\bar{q}^{(t)}_{u,\bl y}}{P(\b Y
=\b y)}.
\label{eq:2}
\end{align}
The recursions above may be implemented by using the matrix
notation, as shown in \cite{bart:06} and \cite{bart:et:al:07a}.

\vspace{-0.1cm}
\section{The class of states selection criteria}
\label{Sec:criteria}
As already discussed in Section \ref{intro}, a crucial point in
using LM models concerns the selection of the number of latent
states $k$. When this number cannot be a priori defined, it is
possible to rely on model selection criteria. In the following we
illustrate the most common log-likelihood based information criteria
together with classification based criteria which take the quality
of classification into account.

\subsection{Log-likelihood based information criteria}
The information criteria are based on indices that are, essentially,
penalized versions of the maximum log-likelihood. The two most
common criteria of this type are AIC and BIC. The first criterion,
proposed by \cite{aka:73}, is a measure of the relative goodness of
fit of a model, which describes the tradeoff between accuracy and
complexity of the model. In particular, AIC is based on estimating
the Kullback-Leibler distance between the true density and the
estimated density, which focuses on the expected log-likelihood, and
is defined on the basis of the following index
\begin{equation}\label{eq:AIC}
\mathrm{AIC} = -2 \;\hat{\ell}(\b\th) + 2 \# \mathrm{par}.
\end{equation}
For a given model, $\hat{\ell}$ denotes the maximum of the
log-likelihood of the LM model of interest and $\# \mathrm{par}$
denotes the number of free parameters as defined in (\ref{eq:par}).
According to this criterion, the optimal number of latent states is
that corresponding to the minimum value of the index in
(\ref{eq:AIC}). In practice, we fit the LM model for increasing
values of $k$ until the index does not start to increase. Then, we
select the previous $k$ as the optimal number of latent states,
which guarantees the best compromise between goodness-of-fit and
model parsimony.

The BIC criterion of \cite{schw:78} is derived, for regular models,
as an approximation to twice the log integrated likelihood
\citep{kass:raftery:1995}, using the Laplace method
\citep{tiern:kada:86}. From the asymptotic behavior of this
approximation, the corresponding index may be defined as
\begin{equation}\label{eq:BIC}
\mathrm{BIC} = -2 \;\hat{\ell}(\b\th) +  \# \mathrm{par}\log(n),
\end{equation}
with $\hat{\ell}$ and $\# \mathrm{par}$ defined as above. In certain
settings, model selection based on BIC is roughly equivalent to
model selection based on Bayes factors; see among others
\cite{kass:raftery:1995}. The number of latent states $k$ to be
selected is the one which corresponds to the minimum value of the
index in (\ref{eq:BIC}). Usually, BIC leads to selecting a smaller
number of latent states than the AIC criterion, since it is based on
a more severe penalization. This difference may be relevant in
complex model. In particular, the BIC criterion is expected to
perform better as the amount of information increases with respect
to the model complexity. In the LM literature, the same criteria
have been used for model selection by \cite{lang:94},
\cite{lang:vand:94}, \cite{magi:verm:01}, among many others.
Finally, a comparison of their performance in connection with states
selection of a
univariate LM model may be found in \cite{bart:farc:penn:13}, Ch.3.
 Moreover, comparisons between AIC and BIC
criteria can be found in the literature of mixture models
\citep[][Ch. 6]{mcla:peel:00}, and in the HM literature for time
series. From these studies, it emerges that BIC is usually
preferable to AIC, as the latter tends to overestimate the number of
states.

Among the variants of the AIC criterion existing in literature we
also consider the criterion introduced by \cite{bozdogan1993}. In
particular, this criterion defines a more penalized version of the
index in (\ref{eq:AIC}), on the basis of the results in
\cite{wolfe:1970}, so as to obtain
\begin{equation}
\mathrm{AIC_3} = -2 \;\hat{\ell}(\b\th) + 3 \# \mathrm{par},
\end{equation}
in which the penalizing term 2 is substituted with 3.

On the other hand, the Consistent AIC criterion (CAIC), proposed by
\cite{Bozdogan1987}, includes a penalizing term which also takes
into account the sample size $n$, and is defined as
\begin{equation}
 \mathrm{CAIC} = -2\; \hat{\ell}(\b\th) + \# \mathrm{par}
 (\log(n)+1).
\end{equation}

Further to the above information criteria that are aimed at
measuring the goodness of fit of a model, we also consider criteria
that take into account the performance of the classification
procedure, as outlined in the following.

\subsection{Classification based information criteria}
The criteria developed in the context of the classification
likelihood approach, also known as complete data information
criteria, are based on data augmentation, that is, the complete data
as defined in Section~\ref{sec:lik}. These criteria consider the
following relation, that was first showed by \cite{hattaway:86}
\begin{equation}\label{eq:ell}
\ell^*(\b \theta) = \ell(\b \theta) - \mathrm{EN},
\end{equation}
see also \cite{cel:sor:96} and \cite{bier:gova:97}, where EN is an
entropy measure, which involves the posterior probabilities of
component membership of each subject belonging to a specific group.
Such entropy may be seen as a penalization term which is a measure
of the ability of the model to provide a relevant partition of the
data. More in detail, if the components are well separated, the
posterior probabilities tend to define a partition of the data,
assuming values close to 1. As a consequence, the entropy will be
close to 0. The entropy measure cannot be directly used to assess
the number of clusters since $\ell(\b \theta)$ is an increasing
function of $k$ and has to be renormalized. With reference to the
mixture models, \cite{cel:sor:96} proposed to consider the NEC
criterion, which is expressed by
\begin{equation}\label{eq:NEC}
\mathrm{NEC} = \frac{\mathrm{EN}}{\hat{\ell}_k(\b \th)- \hat{\ell}_1(\b \th)}, \quad
k\geq 2,
\end{equation}
where $\hat{\ell}_k(\b\th)$ is the maximum log-likelihood in case of
a $k$ components mixture and  $\hat{\ell}_1(\b\th)$ is the maximum
log-likelihood in case of a $1$ component mixture. As also
illustrated in \cite{bier:gova:97}, NEC must assume small values to
obtain a compromise between a good classification feature and a good
description of the data. Then, the optimal number of components is
the one that minimizes the index in (\ref{eq:NEC}). It is worth
noting that the NEC criterion is not defined when $k=1$; to deal
with this problem \cite{bier:et:al:99} proposed an empirical version
of the NEC for Gaussian mixture. 
Usually it is convention to use $\mathrm{NEC}=1$ for $k=1$.

In extending NEC to LM models, the difficulty is in considering
entropy based on the posterior probabilities of all the possible
configurations of latent states, given the observed data, for every
sample unit. Therefore, this ``true'' entropy can be computed only
when we have a reduced number of times occasions and latent states.
In the context of HM models this measure is defined by
\cite{hern:cres:cybe:05} as
\begin{align*}
\mathrm{EN} & = - \sum_{u_1} \ldots \sum_{u_T} f_{u_1, \ldots u_T|\bl y} \log (f_{u_1, \ldots u_T|\bl y}) =
\\
   & = - \sum_{u_1} \ldots \sum_{u_T} f_{u_1|\bl y}^{(1)} \cdot f_{u_2|u_1,\bl y}^{(2|1)} \cdot \ldots  \cdot  f_{u_t|u_{t-1},\bl y}^{(t|t-1)}  \cdot \ldots  \cdot f_{u_T|u_{T-1},\bl y}^{(T|T-1)} \cdot \\
   & \cdot [\log (f_{u_1|\bl y}^{(1)}) +\log (f_{u_2|u_1,\bl y}^{(2|1)})+ \ldots + \log (f_{u_t|u_{t-1},\bl y}^{(t|t-1)})  +\ldots  +\log ( f_{u_T|u_{T-1},\bl y}^{(T|T-1)})     ]
\end{align*}
where $f_{u|\bl y}^{(t)}$ and $f_{u|v, \bl y}^{(t|t-1)} $ are
defined in equations (\ref{eq:1}) and (\ref{eq:2}), respectively.

In the context of LM models, we may simplify the above equation for
EN by formulating an approximated version that allows us to compute
entropy also for any number of time occasions and latent states.
More precisely, under the assumption that $u^{(t)}$ are independent
given $\b Y$, we define EN as follows
\[
\mathrm{EN}_1 = -\sum_{u_1} \ldots \sum_{u_T} f_{u|\bl y}^{(t)} \log
(f_{u|\bl y}^{(t)}).
\]
A possible renormalized variant of  $EN_1$ may also be expressed as
\[
\mathrm{EN}_2 = - \sum_{u_1} \ldots \sum_{u_T} f_{u|\bl y}^{(t)} \log
(f_{u|\bl y}^{(t)})/T.
\]
Therefore, we consider a NEC criterion relying on the ``true''
entropy based on the posterior probabilities of all possible
configurations of latent states, and two different approximated
versions, which may be computed for any number of time occasions and
latent states.
As an example we suppose to observe subjects at three occasions
$(T=3)$, then the above criteria may be explicitly written as follows
 \begin{align*}
   \mathrm{EN} & = - \sum_{u}\sum_{v} \sum_{z} f_{u, v, z | \bl y} \log (f_{u, v, z | \bl y}) = \\
& = f_{z |v,  \bl y}^{(3|2)} \cdot f_{v | u,  \bl y}^{(2|1)} \cdot f_{u |   \bl y}^{(1)}  \cdot\\
& \cdot [\log (f_{z | v,  \bl y}^{(3|2)}) + \log (f_{v| u,  \bl
y}^{(2|1)}) + \log (f_{u |   \bl y}^{(1)}) ] \\ \vspace{5mm}
 \mathrm{EN_1} & =  - [f_{u |  \bl y}^{(1)} \cdot \log (f_{u |   \bl y}^{(1)}) +  f_{v |  \bl y}^{(2)} \cdot  \log (f_{v |  \bl y}^{(2)}) +  f_{z | \bl y}^{(3)} \cdot \log (f_{z |   \bl y}^{(3)})]  \\ \vspace{5mm}
 \mathrm{EN_2} & =   \frac{1}{3}  \mathrm{EN_1}
\end{align*}
According with the entropy measures defined above we consider three
different versions of the NEC criterion where the first one is based
on the ``true'' entropy, as defined in (\ref{eq:NEC}), and the other
two versions are expressed as
\begin{equation*}
\mathrm{NEC}_1 = \frac{\mathrm{EN}_1}{\hat{\ell}(\b\th) - \hat{\ell}_1(\b\th)}, \quad
k \geq 2;
\end{equation*}
\begin{equation*}
\mathrm{NEC}_2 = \frac{\mathrm{EN}_2}{\hat{\ell}(\b\th) - \hat{\ell}_1(\b\th)}, \quad
k \geq 2.
\end{equation*}

Among other criteria which take the quality of classification into
account, we also consider the CLC criterion, proposed by
\cite{bier:gova:97} in the mixture context, which uses the relation
in (\ref{eq:ell}) to define the following index
\begin{equation*}
\mathrm{CLC} = -2 \hat{\ell}(\b\th) + 2 \; \textrm{EN}.
\end{equation*}

Moreover, \cite{bier:cele:gova:98} suggested an alternative
information criterion based on the complete data likelihood named as
Integrated Classification Likelihood criterion (ICL). The same
authors also proposed an approximated version of the ICL using BIC
\citep{bier:cele:gova:00}. In particular, \cite{mcla:peel:00}
referred to this approximated version as ICL-BIC and showed that may
be computed as
\begin{equation*}
\textrm{ICL-BIC} = \textrm{BIC}  + 2 \; \textrm{EN},
\end{equation*}
in which the term $2\; \textrm{EN}$ represents a kind of penalization for
poorly separated clusters; see also \cite{li:05}.

As already discussed in Section~\ref{intro}, although the above
information criteria are widely used and their performance are
studied in the context of finite mixture, LC, and HM models,
there is still a lack in the literature about their comparison in
the context of LM models. In the following section we set an
experimental design to compare the performance of these criteria in
the context of multivariate LM model, in order to choose the optimal
number of latent states.

\vspace{-0.1cm}
\section{Simulation study}
\label{Sec:sim}

We illustrate the results obtained by a Monte Carlo simulation study
aimed at comparing the performance of the following indices for
states selection:
\begin{itemize}
    \item Log-likelihood based criteria: AIC, CAIC, AIC$_3$, BIC;
    \item Classification based criteria: CLC, ICL-BIC, NEC, NEC$_1$, NEC$_2$.
\end{itemize}
More in detail, we simulate 100 samples with a given size $n$
($n=250,500$) and coming from an LM model, characterized by $r$
($r=1,3,5$) binary ($y=0,1$) response variables observed in $T=5$
time occasions, $k$ ($k=2,3$) latent states, and given values of
initial probabilities $\pi_u$, transition probabilities
$\pi_{u|v}^{(t|t-1)}$, and conditional response probabilities
$\phi_{jy|u}^{(t)}$. All analyses are implemented in
\texttt{R} software (the code is available upon request by authors).
In all cases, the strategy adopted to choose the number of latent
states is the same: we fit the LM model with increasing $k$ values
and, then, the value just before the first increasing criterion
index is taken as optimal number of latent states.

We first consider a scenery (scenery 1) based on $n=250$ individuals
belonging to $k=2$ latent states and observed on $T=5$ time
occasions. Moreover, we suppose equal initial probabilities, that is
$\pi_1=0.5=\pi_2$, and denoting  by $ \b \Pi$ the transition
probabilities matrix with
%
elements  $\pi_{u|v}$, under the time homogeneous assumption, we
consider
\[
\bPi =  \begin{array}{ll}
\left(\begin{matrix}
0.9 &  0.1 \\
0.1 & 0.9
\end{matrix}\right).
\end{array}
\]
We also assume alternatively $r= 1,3,5$ binary response variables
with the following matrix $\bPhi$ of the conditional response
probabilities $\phi_{jy|u}^{(t)}$
\[
\bPhi =  \begin{array}{ll}
\left(\begin{matrix}
0.8 &  0.2 \\
0.2 & 0.8
\end{matrix}\right).
\end{array}
\]

Table \ref{tab:0} shows the relative frequencies of the number $k$ of components chosen by each of the considered criteria, both in univariate case ($r=1$) and in multivariate cases ($r=3$ and $r=5$).

\begin{table}[!ht]\vspace{-1.5mm}
\caption{Relative frequencies of $k$ chosen on the basis of several
criteria (scenery 1, $n=250$)} \label{tab:0}
\begin{tabular}{llllllllll}
\hline\noalign{\smallskip}
$k$ &   BIC &   AIC &   AIC$_3$    &   CAIC    &   NEC &   NEC$_1$ &   NEC$_2$ &   CLC &   ICL-BIC \\
\noalign{\smallskip}\hline\noalign{\smallskip}
  $r=1$  &      &       &       &       &       &       &       &       &   \\
1   &   0.00    &   0.00    &   0.00    &   0.00    &   \textbf{1.00}   &   \textbf{0.95}   &   \textbf{1.00}   &   \textbf{1.00}   &    \textbf{1.00}  \\
2   &   \textbf{1.00}   &   \textbf{0.99}   &   \textbf{1.00}   &   \textbf{1.00}   &   0.00    &   0.05    &   0.00    &   0.00    &   0.00    \\
3   &   0.00    &   0.01    &   0.00    &   0.00    &   0.00    &   0.00    &   0.00    &   0.00    &   0.00    \\
4   &   0.00    &   0.00    &   0.00    &   0.00    &   0.00    &   0.00    &   0.00    &   0.00    &   0.00    \\
5   &   0.00    &   0.00    &   0.00    &   0.00    &   0.00    &   0.00    &   0.00    &   0.00    &   0.00    \\
$r=3$  &      &       &       &       &       &       &       &       &   \\
1   &   0.00    &   0.00    &   0.00    &   0.00    &   0.00    &   0.00    &   0.00    &   0.00    &   0.00    \\
2   &   \textbf{1.00}   &   \textbf{0.83}   &   \textbf{1.00}   &   \textbf{1.00}   &   \textbf{0.99}   &   \textbf{1.00}   &   \textbf{1.00}       &   \textbf{0.98}   &   \textbf{1.00}   \\
3   &   0.00    &   0.14    &   0.00    &   0.00    &   0.00    &   0.00    &   0.01    &   0.00    &   0.00    \\
4   &   0.00    &   0.03    &   0.00    &   0.00    &   0.00    &   0.00    &   0.00    &   0.00    &   0.00    \\
5   &   0.00    &   0.00    &   0.00    &   0.01    &   0.00    &   0.00    &   0.01    &   0.00    &   0.00    \\
$r=5$  &      &       &       &       &       &       &       &       &   \\
1   &   0.00    &   0.00    &   0.00    &   0.00    &   0.00    &   0.00    &   0.00    &   0.00    &   0.00    \\
2   &   \textbf{1.00}   &   \textbf{0.71}   &   \textbf{0.97}   &   \textbf{1.00}   &   \textbf{0.99}   &   \textbf{1.00}   &   \textbf{1.00}   &   \textbf{0.93}   &   \textbf{1.00}   \\
3   &   0.00    &   0.26    &   0.03    &   0.00    &   0.01    &   0.00    &   0.00    &   0.05    &   0.00    \\
4   &   0.00    &   0.03    &   0.00    &   0.00    &   0.00    &   0.00    &   0.00    &   0.00    &   0.00    \\
5   &   0.00    &   0.00    &   0.00    &   0.00    &   0.00    &   0.00    &   0.00    &   0.02    &   0.00    \\
\noalign{\smallskip}\hline
\end{tabular}\vspace{-1.5mm}
\end{table}

In the univariate case, all log-likelihood based criteria perform
very well, whereas the performance of classification based criteria
is very bad: they tend to underestimate $k$ in almost all cases.
Instead, in the multivariate cases ($r=3$ and $r=5$) the
classification based criteria improve considerably their
performance, being the selected number of latent states equal to 2
in almost all cases. We only  observe a worsening of AIC, which
tends to overestimate the right number of latent states in 17 and 29
cases out of 100, for $r=3$ and for $r=5$ respectively.

An alternative scenery (scenery 2) is then considered, which differs from scenery 1 for lower values of state persistence probabilities given by
\[
\bPi =  \begin{array}{ll}
\left(\begin{matrix}
0.7 &  0.3 \\
0.3 & 0.7
\end{matrix}\right).
\end{array}
\]
All the other elements are the same than those considered in scenery 1. Results are shown in Table \ref{tab:1}.

\begin{table}[!ht]\vspace{-1.5mm}
\caption{Relative frequencies of $k$ chosen on the basis of several
criteria (scenery 2, $n=250$)} \label{tab:1}
\begin{tabular}{llllllllll}
\hline\noalign{\smallskip}
$k$ &   BIC &   AIC &   AIC$_3$   &   CAIC    &   NEC &   NEC$_1$ &   NEC$_2$ &   CLC &   ICL-BIC \\
\noalign{\smallskip}\hline\noalign{\smallskip}
  $r=1$  &      &       &       &       &       &       &       &       &   \\
1   &   \textbf{0.52}    &   0.00    &   0.10    &   \textbf{0.63}    &   \textbf{1.00}    &   \textbf{1.00}    &   \textbf{0.99}    &   \textbf{1.00}    &   \textbf{1.00}    \\
2   &   0.48    &   \textbf{0.98}    &   \textbf{0.90}    &   0.37    &   0.00    &   0.00    &   0.01    &   0.00    &   0.00    \\
3   &   0.00    &   0.02    &   0.00    &   0.00    &   0.00    &   0.00    &   0.00    &   0.00    &   0.00    \\
4   &   0.00    &   0.00    &   0.00    &   0.00    &   0.00    &   0.00    &   0.00    &   0.00    &   0.00    \\
5   &   0.00    &   0.00    &   0.00    &   0.00    &   0.00    &   0.00    &   0.00    &   0.00    &   0.00    \\
$r=3$   &       &       &       &       &       &       &       &       &   \\
1   &   0.00    &   0.00    &   0.00    &   0.00    &   \textbf{0.88}    &   \textbf{0.92}    &   0.00    &   \textbf{0.88}    &   \textbf{0.95}    \\
2   &   \textbf{1.00}    &   \textbf{0.83}    &   \textbf{0.98}    &   \textbf{1.00}    &   0.10    &   0.07    &   \textbf{0.96}    &   0.10    &   0.04    \\
3   &   0.00    &   0.16    &   0.02    &   0.00    &   0.01    &   0.01    &   0.04    &   0.01    &   0.01    \\
4   &   0.00    &   0.01    &   0.00    &   0.00    &   0.01    &   0.00    &   0.00    &   0.01    &   0.00    \\
5   &   0.00    &   0.00    &   0.00    &   0.00    &   0.00    &   0.00    &   0.00    &   0.00    &   0.00    \\
$r=5$   &       &       &       &       &       &       &       &       &   \\
1   &   0.00    &   0.00    &   0.00    &   0.00    &   0.00    &   0.00    &   0.00    &   0.00    &   0.00    \\
2   &   \textbf{1.00}    &   \textbf{0.77}    &   \textbf{1.00}    &   \textbf{1.00}    & \textbf{1.00}  &   \textbf{1.00}    &   \textbf{1.00}    &   \textbf{1.00}    &  \textbf{1.00} \\
3   &   0.00    &   0.15    &   0.00    &   0.00    &   0.00    &   0.00    &   0.00    &   0.00    &   0.00    \\
4   &   0.00    &   0.06    &   0.00    &   0.00    &   0.00    &   0.00    &   0.00    &   0.00    &   0.00    \\
5   &   0.00    &   0.02    &   0.00    &   0.00    &   0.00    &   0.00    &   0.00    &   0.00    &   0.00    \\
\noalign{\smallskip}\hline
\end{tabular}\vspace{-1.5mm}
\end{table}

With respect to scenery 1 we note several differences. In the
univariate case, the behavior of both BIC and CAIC gets worse: BIC
leads to select the true value $k=2$ in less than 50$\%$ of cases
and CAIC in 37$\%$ of cases, whereas in the remaining simulations
both of them underestimate $k$. A slightly worsening is observed
also for AIC$_3$, which in the 10$\%$ of simulated models chooses
only one latent state. With $r=3$ responses, on one hand BIC,
AIC$_3$, CAIC and NEC$_2$ considerably  improve their performance
and, on the other hand, AIC tends to overestimate $k$ in 17 cases
out of 100, obtaining in both situations values similar to those of
scenery 1. However, the classification based criteria other than
NEC$_2$ improve just a little and continue  to underestimate $k$ in
the main part of cases, showing a really different behavior with
respect to scenery 1. Similarly to scenery 1, with $r=5$ all the
considered criteria present an optimal behavior (with the exception
of AIC, which overestimates $k$ in 23 cases out of 100).

Another scenery (scenery 3) is then considered, which differs from scenery 1 for a greater uncertainty in the allocation of the observations to the latent states, being the  conditional response probabilities matrix given by
\[
\bPhi =  \begin{array}{ll}
\left(\begin{matrix}
0.7 &  0.3 \\
0.3 & 0.7
\end{matrix}\right).
\end{array}
\]
All the other elements are the same than scenery 1. Results are shown in Table \ref{tab:2}.

\begin{table}[!ht]
\caption{Relative frequencies of $k$ chosen on the basis of several
criteria (scenery 3, $n=250$)} \label{tab:2}
\begin{tabular}{llllllllll}
\hline\noalign{\smallskip}
$k$ &   BIC &   AIC &   AIC$_3$    &   CAIC    &   NEC &   NEC$_1$ &   NEC$_2$ &   CLC &   ICL-BIC \\
\noalign{\smallskip}\hline\noalign{\smallskip}
  $r=1$  &      &       &       &       &       &       &       &       &   \\
1   &   0.35    &   0.01    &   0.02    &   \textbf{0.53}    &   \textbf{1.00}    &   \textbf{1.00}    &   \textbf{1.00}    &   \textbf{1.00}   &   \textbf{1.00}    \\
2   &   \textbf{0.65}    &   \textbf{0.98}    &   \textbf{0.97}    &   0.47    &   0.00    &   0.00    &   0.00    &   0.00    &   0.00    \\
3   &   0.00    &   0.01    &   0.01    &   0.00    &   0.00    &   0.00    &   0.00    &   0.00    &   0.00    \\
4   &   0.00    &   0.00    &   0.00    &   0.00    &   0.00    &   0.00    &   0.00    &   0.00    &   0.00    \\
5   &   0.00    &   0.00    &   0.00    &   0.00    &   0.00    &   0.00    &   0.00    &   0.00    &   0.00    \\
$r=3$   &       &       &       &       &       &       &       &       &   \\
1   &   0.00    &   0.00    &   0.00    &   0.00    &   \textbf{0.99}    &   \textbf{1.00}    &   0.08   &   \textbf{0.99}    &   \textbf{1.00}    \\
2   &   \textbf{1.00}    &   \textbf{0.79}    &   \textbf{1.00}   &   \textbf{1.00}    &   0.01    &   0.00    &   \textbf{0.88}   &   0.01    &   0.00    \\
3   &   0.00    &   0.18    &   0.00   &   0.00    &   0.00    &   0.00    &   0.03   &   0.00    &   0.00    \\
4   &   0.00    &   0.03    &   0.00    &   0.00    &   0.00    &   0.00    &   0.00   &   0.00    &   0.00    \\
5   &   0.00    &   0.00    &   0.00    &   0.00    &   0.00    &   0.00    &   0.01   &   0.00    &   0.00    \\
$r=5$   &       &       &       &       &       &       &       &       &   \\
1   &   0.00    &   0.00    &   0.000   &   0.00    &   0.285   &   \textbf{0.770}   &   0.000   &   0.285   &   \textbf{0.550}   \\
2   &   \textbf{1.00}    &   \textbf{0.78}    &   \textbf{0.995}   &  \textbf{1.00} &   \textbf{0.590}   &   0.220   &   \textbf{0.980}   &   \textbf{0.585} & 0.445   \\
3   &   0.00    &   0.205   &   0.005   &   0.00    &   0.030   &   0.005   &   0.015   &   0.035   &   0.005   \\
4   &   0.00    &   0.01    &   0.00    &   0.00    &   0.070   &   0.005   &   0.005   &   0.070   &   0.000   \\
5   &   0.00    &   0.005   &   0.00    &   0.00    &   0.025   &   0.000   &   0.000   &   0.025   &   0.000   \\
\noalign{\smallskip}\hline
\end{tabular}\vspace{-1.5mm}
\end{table}

With respect to scenery 1, in presence of $r=1$ response variables
the behavior of BIC and CAIC is not very satisfactory, because they
tend to underestimate $k$ in 35$\%$ and 53$\%$ of cases. Concerning
the classification based criteria, we note a significant
deterioration of their behavior  both in univariate case and in
multivariate cases.  Only NEC$_2$ presents a satisfactory
performance, being the correct number of $k$ selected in 88 cases
out of 100 with $r=3$ and in 98 cases out of 100 with $r=5$ (and
overestimated in the remaining cases). Instead, the remaining
criteria lead to choose $k=1$ in almost all cases when $r=3$,
improving just a little when $r=5$. More precisely, in this last
case, NEC and CLC allow us to select the right number of $k$ in the
59$\%$ of cases, whereas they underestimate $k$ in $28.5\%$ of
cases. 
Moreover, ICL-BIC
leads to select $k=1$ for the 55$\%$ of simulated models and NEC$_1$
for the 77$\%$ of them.

The three above described sceneries are then replicated by increasing the number of observations from $n=250$ to $n=500$, all the other things being constant. Results are shown in Tables \ref{tab:0a}, \ref{tab:1a}, and \ref{tab:2a} for sceneries 1, 2, and 3 respectively.

\begin{table}[!ht]\vspace{-1.5mm}
\caption{Relative frequencies of $k$ chosen on the basis of several
criteria (scenery 1, $n=500$)} \label{tab:0a}
\begin{tabular}{llllllllll}
\hline\noalign{\smallskip}
$k$ &   BIC &   AIC &   AIC$_3$    &   CAIC    &   NEC &   NEC$_1$ &   NEC$_2$ &   CLC &   ICL-BIC \\
\noalign{\smallskip}\hline\noalign{\smallskip}
  $r=1$  &      &       &       &       &       &       &       &       &   \\
1   &   0.00    &   0.00    &   0.00    &   0.00    &   \textbf{1.00}   &   \textbf{1.00}   &   \textbf{0.98}   &   \textbf{1.00}   &   \textbf{1.00}   \\
2   &   \textbf{1.00}   &   \textbf{0.99}   &   \textbf{1.00}   &   \textbf{1.00}   &   0.00    &   0.00    &   0.02    &   0.00    &   0.00    \\
3   &   0.00    &   0.01    &   0.00    &   0.00    &   0.00    &   0.00    &   0.00    &   0.00    &   0.00    \\
4   &   0.00    &   0.00    &   0.00    &   0.00    &   0.00    &   0.00    &   0.00    &   0.00    &   0.00    \\
5   &   0.00    &   0.00    &   0.00    &   0.00    &   0.00    &   0.00    &   0.00    &   0.00    &   0.00    \\
 $r=3$  &       &       &       &       &       &       &       &       &       \\
1   &   0.00    &   0.00    &   0.00    &   0.00    &   0.00    &   0.00    &   0.00    &   0.00    &   0.00    \\
2   &   \textbf{1.00}   &   \textbf{0.87}   &   \textbf{1.00}   &   \textbf{1.00}   &   \textbf{0.99}   &   \textbf{1.00}   &   \textbf{1.00}   &   \textbf{0.99}   &   \textbf{1.00}   \\
3   &   0.00    &   0.13    &   0.00    &   0.00    &   0.01    &   0.00    &   0.00    &   0.01    &   0.00    \\
4   &   0.00    &   0.00    &   0.00    &   0.00    &   0.00    &   0.00    &   0.00    &   0.00    &   0.00    \\
5   &   0.00    &   0.00    &   0.00    &   0.00    &   0.00    &   0.00    &   0.00    &   0.00    &   0.00    \\
 $r=5$  &       &       &       &       &       &       &       &       &       \\
1   &   0.00    &   0.00    &   0.00    &   0.00    &   0.00    &   0.00    &   0.00    &   0.00    &   0.00    \\
2   &   \textbf{1.00}   &   \textbf{0.78}   &   \textbf{1.00}   &   \textbf{1.00}   &   \textbf{1.00}   &   \textbf{1.00}   &   \textbf{1.00}   &   \textbf{1.00}   &   \textbf{1.00}   \\
3   &   0.00    &   0.17    &   0.00    &   0.00    &   0.00    &   0.00    &   0.00    &   0.00    &   0.00    \\
4   &   0.00    &   0.05    &   0.00    &   0.00    &   0.00    &   0.00    &   0.00    &   0.00    &   0.00    \\
5   &   0.00    &   0.00    &   0.00    &   0.00    &   0.00    &   0.00    &   0.00    &   0.00    &   0.00    \\
\noalign{\smallskip}\hline
\end{tabular}
\end{table}
\begin{table}[!ht]
\caption{Relative frequencies of $k$ chosen on the basis of several
criteria (scenery 2, $n=500$)} \label{tab:1a}
\begin{tabular}{llllllllll}
\hline\noalign{\smallskip}
$k$ &   BIC &   AIC &   AIC$_3$    &   CAIC    &   NEC &   NEC$_1$ &   NEC$_2$ &   CLC &   ICL-BIC \\
\noalign{\smallskip}\hline\noalign{\smallskip}
  $r=1$  &      &       &       &       &       &       &       &       &   \\
1   &   0.05    &   0.00    &   0.01    &   0.08    &   \textbf{1.00}   &   \textbf{1.00}   &   \textbf{1.00}   &   \textbf{1.00}   &   \textbf{1.00}   \\
2   &   \textbf{0.95}   &   \textbf{0.99}   &   \textbf{0.99}   &   \textbf{0.92}   &   0.00    &   0.00    &   0.00    &   0.00    &   0.00    \\
3   &   0.00    &   0.01    &   0.00    &   0.00    &   0.00    &   0.00    &   0.00    &   0.00    &   0.00    \\
4   &   0.00    &   0.00    &   0.00    &   0.00    &   0.00    &   0.00    &   0.00    &   0.00    &   0.00    \\
5   &   0.00    &   0.00    &   0.00    &   0.00    &   0.00    &   0.00    &   0.00    &   0.00    &   0.00    \\
 $r=3$  &       &       &       &       &       &       &       &       &       \\
1   &   0.00    &   0.00    &   0.00    &   0.00    &   \textbf{0.97}   &   \textbf{0.98}   &   0.00    &   \textbf{0.97}   &   \textbf{0.98}   \\
2   &   \textbf{1.00}   &   \textbf{0.74}   &   \textbf{0.99}   &   \textbf{1.00}   &   0.03    &   0.02    &   \textbf{1.00}   &   0.03    &   0.02    \\
3   &   0.00    &   0.23    &   0.01    &   0.00    &   0.00    &   0.00    &   0.00    &   0.00    &   0.00    \\
4   &   0.00    &   0.02    &   0.00    &   0.00    &   0.00    &   0.00    &   0.00    &   0.00    &   0.00    \\
5   &   0.00    &   0.01    &   0.00    &   0.00    &   0.00    &   0.00    &   0.00    &   0.00    &   0.00    \\
 $r=5$  &       &       &       &       &       &       &       &       &       \\
1   &   0.00    &   0.00    &   0.00    &   0.00    &   0.00    &   0.00    &   0.00    &   0.00    &   0.00    \\
2   &   \textbf{1.00}   &   \textbf{0.74}   &   \textbf{0.98}   &   \textbf{1.00}   &   \textbf{1.00}   &   \textbf{1.00}   &   \textbf{1.00}   &   \textbf{1.00}   &   \textbf{1.00}   \\
3   &   0.00    &   0.17    &   0.02    &   0.00    &   0.00    &   0.00    &   0.00    &   0.00    &   0.00    \\
4   &   0.00    &   0.08    &   0.00    &   0.00    &   0.00    &   0.00    &   0.00    &   0.00    &   0.00    \\
5   &   0.00    &   0.01    &   0.00    &   0.00    &   0.00    &   0.00    &   0.00    &   0.00    &   0.00    \\
\noalign{\smallskip}\hline
\end{tabular}\vspace{-1.5mm}
\end{table}

\begin{table}[!ht]
\caption{Relative frequencies of $k$ chosen on the basis of several criteria (scenery 3, $n=500$)}
\label{tab:2a}
\begin{tabular}{llllllllll}
\hline\noalign{\smallskip}
$k$ &   BIC &   AIC &   AIC$_3$    &   CAIC    &   NEC &   NEC$_1$ &   NEC$_2$ &   CLC &   ICL-BIC \\
\noalign{\smallskip}\hline\noalign{\smallskip}
  $r=1$  &      &       &       &       &       &       &       &       &   \\
1   &   0.01    &   0.00    &   0.00    &   0.03    &   \textbf{1.00}   &   \textbf{1.00}   &   \textbf{1.00}   &   \textbf{1.00}   &   \textbf{1.00}   \\
2   &   \textbf{0.99}   &   \textbf{1.00}   &   \textbf{1.00}   &   \textbf{0.97}   &   0.00    &   0.00    &   0.00    &   0.00    &   0.00    \\
3   &   0.00    &   0.00    &   0.00    &   0.00    &   0.00    &   0.00    &   0.00    &   0.00    &   0.00    \\
4   &   0.00    &   0.00    &   0.00    &   0.00    &   0.00    &   0.00    &   0.00    &   0.00    &   0.00    \\
5   &   0.00    &   0.00    &   0.00    &   0.00    &   0.00    &   0.00    &   0.00    &   0.00    &   0.00    \\
$r=3$   &       &       &       &       &       &       &       &       &       \\
1   &   0.00    &   0.00    &   0.00    &   0.00    &   \textbf{1.00}   &   \textbf{1.00}   &   0.04    &   \textbf{1.00}   &   \textbf{1.00}   \\
2   &   \textbf{1.00}   &   \textbf{0.86}   &   \textbf{0.97}   &   \textbf{1.00}   &   0.00    &   0.00    &   \textbf{0.96}   &   0.00    &   0.00    \\
3   &   0.00    &   0.12    &   0.03    &   0.00    &   0.00    &   0.00    &   0.000   &   0.00    &   0.00    \\
4   &   0.00    &   0.01    &   0.00    &   0.00    &   0.00    &   0.00    &   0.000   &   0.00    &   0.00    \\
5   &   0.00    &   0.01    &   0.00    &   0.00    &   0.00    &   0.00    &   0.000   &   0.00    &   0.00    \\
$r=5$   &       &       &       &       &       &       &       &       &       \\
1   &   0.00    &   0.00    &   0.00    &   0.00    &   0.32    &   \textbf{0.90}   &   0.00    &   0.32    &   \textbf{0.57}   \\
2   &   \textbf{1.00}   &   \textbf{0.67}   &   \textbf{1.00}   &   \textbf{1.00}   &   \textbf{0.63}   &   0.10    &   \textbf{1.00}   &   \textbf{0.63}   &   0.43    \\
3   &   0.00    &   0.27    &   0.00    &   0.00    &   0.01    &   0.00    &   0.00    &   0.01    &   0.00    \\
4   &   0.00    &   0.04    &   0.00    &   0.00    &   0.03    &   0.00    &   0.00    &   0.03    &   0.00    \\
5   &   0.00    &   0.02    &   0.00    &   0.00    &   0.01    &   0.00    &   0.00    &   0.01    &   0.00    \\
\noalign{\smallskip}\hline
\end{tabular}\vspace{-1.5mm}
\end{table}

By increasing the number of observations we note a considerable
improvement of performances of BIC and CAIC in the univariate cases
of sceneries 2 and 3, whereas the behavior of the other criteria,
especially of classification based criteria, is unchanged. Rather,
in case of scenery 3, when $r=5$ response variables are considered,
the behavior of NEC$_1$ and ICL-BIC gets worse, being $k=2$ selected
in 10$\%$ and 43$\%$ of cases for $n=500$ against 22$\%$ and
44.5$\%$ of cases for $n=250$.

To conclude, we also consider two further sceneries (sceneries 4 and 5) characterized by $n=500$ individuals and  $k=3$ latent states. We also suppose $T=5$, equal initial probabilities, that is $\pi_1=\pi_2=\pi_3=1/3$,
conditional response probabilities matrix equal to
\[
 \begin{array}{lll}
\bPhi = \left(\begin{matrix}
0.9 &  0.1& 0.7 \\
0.1 & 0.9 & 0.3 \\
\end{matrix}\right),
\end{array}
\]
and the following transition probabilities (under the time homogeneous assumption)
\[
 \begin{array}{lll}
\bPi = \left(\begin{matrix}
0.90 &  0.05 & 0.05 \\
0.05 & 0.90 & 0.05 \\
0.05 & 0.05 & 0.90 \\
\end{matrix}\right),
\end{array}
\]
in the first case (scenery 4) and
\[
 \begin{array}{lll}
\bPi= \left(\begin{matrix}
0.70 &  0.15 & 0.15 \\
0.15 & 0.70 & 0.15 \\
0.15 & 0.15 & 0.70 \\
\end{matrix}\right),
\end{array}
\]
in the second case (scenery 5).

With respect to the cases with $k=2$ latent states, we now observe a very poor performance of all criteria in the univariate case ($r=1$): log-likelihood based criteria lead to select $k=2$ and classification based criteria lead to select $k=1$, in almost all cases (Tables \ref{tab:3} and \ref{tab:3a}). Instead, in presence of $r=3$ response variables, the behavior of log-likelihood based criteria gets better, especially under scenery 4. Indeed, Table \ref{tab:3} shows that with AIC and AIC$_3$ the right number of latent states is chosen in 91 and 98 cases out of 100 respectively, whereas with BIC and CAIC the percentages of right choices are reduced to 59$\%$ and $39\%$ respectively, being $k$ underestimated in the remaining cases. On the other hand, under scenery 5 (Table \ref{tab:3a}), which refers to a situation with lower latent states persistence probabilities, the improvement is far from clear: $k=3$ is selected by AIC in 75$\%$ of cases and by AIC$_3$ in 35$\%$, whereas BIC and CAIC lead to choose regularly $k=2$.  Finally, with $r=5$ responses we observe satisfactory performances of log-likelihood based criteria (although AIC overestimates $k$ in 15$\%$ of cases) in case of scenery 4 (Table \ref{tab:3}), but, again under scenery 5, results are not very satisfactory, because BIC and CAIC performs well in only 76$\%$ and 52$\%$ of cases.
The behavior of classification based criteria is definitely disappointing under both sceneries and both with $r=3$ and $r=5$.

\begin{table}[!ht]
\caption{Relative frequencies of $k$ chosen on the basis of several criteria (scenery 4)}
\label{tab:3}
\begin{tabular}{llllllllll}
\hline\noalign{\smallskip}
$k$ &   BIC &   AIC &   AIC$_3$    &   CAIC    &   NEC &   NEC$_1$ &   NEC$_2$ &   CLC &   ICL-BIC \\
\noalign{\smallskip}\hline\noalign{\smallskip}
 $r=1$  &      &       &       &       &       &       &       &       &   \\
1   &   0.00    &   0.00    &   0.00    &   0.00    &   \textbf{1.00}   &   \textbf{1.00}   &   0.00    &   \textbf{1.00}   &   \textbf{1.00}   \\
2   &   \textbf{1.00}   &   \textbf{0.93}   &   \textbf{0.99}   &   \textbf{1.00}   &   0.00    &   0.00    &   \textbf{1.00}   &   0.00    &   0.00    \\
3   &   0.00    &   0.07    &   0.01    &   0.00    &   0.00    &   0.00    &   0.00    &   0.00    &   0.00    \\
4   &   0.00    &   0.00    &   0.00    &   0.00    &   0.00    &   0.00    &   0.00    &   0.00    &   0.00    \\
5   &   0.00    &   0.00    &   0.00    &   0.00    &   0.00    &   0.00    &   0.00    &   0.00    &   0.00    \\
  $r=3$     &       &       &       &       &       &       &       &       &       \\
1   &   0.00    &   0.00    &   0.00    &   0.00    &   0.00    &   0.00    &   0.00    &   0.00    &   0.00    \\
2   &   0.41    &   0.01    &   0.02    &   \textbf{0.62}   &   \textbf{1.00}   &   \textbf{1.00}   &   \textbf{1.00}   &   \textbf{1.00}   &   \textbf{1.00}   \\
3   &   \textbf{0.59}   &   \textbf{0.91}   &   \textbf{0.98}   &   0.39    &   0.00    &   0.00    &   0.00    &   0.00    &   0.00    \\
4   &   0.00    &   0.07    &   0.00    &   0.00    &   0.00    &   0.00    &   0.00    &   0.00    &   0.00    \\
5   &   0.00    &   0.01    &   0.00    &   0.00    &   0.00    &   0.00    &   0.00    &   0.00    &   0.00    \\
  $r=5$     &       &       &       &       &       &       &       &       &       \\
1   &   0.00    &   0.00    &   0.00    &   0.00    &   0.00    &   0.00    &   0.00    &   0.00    &   0.00    \\
2   &   0.00    &   0.00    &   0.00    &   0.00    &   \textbf{1.00}   &   \textbf{1.00}   &   \textbf{1.00}   &   \textbf{1.00}   &   \textbf{1.00}   \\
3   &   \textbf{1.00}   &   \textbf{0.85}   &   \textbf{0.99}   &   \textbf{1.00}   &   0.00    &   0.00    &   0.00    &   0.00    &   0.00    \\
4   &   0.00    &   0.13    &   0.01    &   0.00    &   0.00    &   0.00    &   0.00    &   0.00    &   0.00    \\
5   &   0.00    &   0.02    &   0.00    &   0.00    &   0.00    &   0.00    &   0.00    &   0.00    &   0.00    \\
\noalign{\smallskip}\hline
\end{tabular}
\end{table}

\begin{table}[!ht]
\caption{Relative frequencies of $k$ chosen on the basis of several criteria (scenery 5)}
\label{tab:3a}
\begin{tabular}{llllllllll}
\hline\noalign{\smallskip}
$k$ &   BIC &   AIC &   AIC$_3$    &   CAIC    &   NEC &   NEC$_1$ &   NEC$_2$ &   CLC &   ICL-BIC \\
\noalign{\smallskip}\hline\noalign{\smallskip}
  $r=1$  &      &       &       &       &       &       &       &       &   \\
1   &   0.00    &   0.00    &   0.00    &   0.00    &   \textbf{1.00}   &   \textbf{1.00}   &   \textbf{1.00}   &   \textbf{1.00}   &   \textbf{1.00}   \\
2   &   \textbf{1.00}   &   \textbf{0.97}   &   \textbf{1.00}   &   \textbf{1.00}   &   0.00    &   0.00    &   0.00    &   0.00    &   0.00    \\
3   &   0.00    &   0.03    &   0.00    &   0.00    &   0.00    &   0.00    &   0.00    &   0.00    &   0.00    \\
4   &   0.00    &   0.00    &   0.00    &   0.00    &   0.00    &   0.00    &   0.00    &   0.00    &   0.00    \\
5   &   0.00    &   0.00    &   0.00    &   0.00    &   0.00    &   0.00    &   0.00    &   0.00    &   0.00    \\
  $r=3$ &       &       &       &       &       &       &       &       &       \\
1   &   0.00    &   0.00    &   0.00    &   0.00    &   0.00    &   0.00    &   0.00    &   0.00    &   0.00    \\
2   &   \textbf{0.99}   &   0.15    &   \textbf{0.65}   &   \textbf{1.00}   &   \textbf{1.00}   &   \textbf{1.00}   &   \textbf{1.00}   &   \textbf{1.00}   &   \textbf{1.00}   \\
3   &   0.01    &   \textbf{0.75}   &   0.35    &   0.00    &   0.00    &   0.00    &   0.00    &   0.00    &   0.00    \\
4   &   0.00    &   0.10    &   0.00    &   0.00    &   0.00    &   0.00    &   0.00    &   0.00    &   0.00    \\
5   &   0.00    &   0.00    &   0.00    &   0.00    &   0.00    &   0.00    &   0.00    &   0.00    &   0.00    \\
  $r=5$ &       &       &       &       &       &       &       &       &       \\
1   &   0.00    &   0.00    &   0.00    &   0.00    &   0.00    &   0.00    &   0.00    &   0.00    &   0.00    \\
2   &   0.24    &   0.00    &   0.01    &   0.48    &   \textbf{1.00}   &   \textbf{1.00}   &   \textbf{1.00}   &   \textbf{1.00}   &   \textbf{1.00}   \\
3   &   \textbf{0.76}   &   \textbf{0.86}   &   \textbf{0.99}   &   \textbf{0.52}   &   0.00    &   0.00    &   0.00    &   0.00    &   0.00    \\
4   &   0.00    &   0.13    &   0.00    &   0.00    &   0.00    &   0.00    &   0.00    &   0.00    &   0.00    \\
5   &   0.00    &   0.01    &   0.00    &   0.00    &   0.00    &   0.00    &   0.00    &   0.00    &   0.00    \\
\noalign{\smallskip}\hline
\end{tabular}\vspace{-2mm}
\end{table}

\vspace{-1cm}
\section{Conclusions}\label{sec:concl}
In this paper we investigated about  a typical issue characterizing
some latent variable models, such as latent class models or hidden
Markov (HM) models, consisting in the choice about the number of
mixture components (i.e, latent classes or latent states). More
precisely, we focused on the selection of latent states in
univariate and multivariate latent Markov (LM) models for
longitudinal data. We firstly illustrated the assumptions and the
structure of LM model, giving some hints about the maximization of
log-likelihood on the basis of an EM algorithm. Then, we described
some of the most well-known model selection criteria used in the
context of mixture models, distinguishing between log-likelihood
based criteria (i.e., AIC, AIC$_3$, CAIC, BIC) and classification
based criteria (NEC, CLC, ICL-BIC). Concerning this latter type of
criteria, we gave some emphasis to the problem of properly defining
the entropy in case of LM models. Relying on the case of HM models,
we observed the possibility of computing the exact entropy only in
presence of a reduced number of time occasions and latent states.
Therefore, we also proposed two variants (named NEC$_1$ and NEC$_2$)
that can be easily computed also for any number of time occasions
and latent states.

On the basis of some Monte Carlo simulations, we compared the
performance of log-likelihood and classification based criteria for
the latent states selection in univariate and multivariate LM
models. Generally speaking, lower values of persistence
probabilities in a same state and/or a greater uncertainty in the
allocation of the observations to the latent states complicate the
task of latent states selection procedures, leading to a generally
worse performance of the adopted criteria. Instead, the number of
observations does not play a relevant role.

Concerning the specific criteria, we observed that those based on
the log-likelihood present a better general behavior with respect to
those based on  classification, even if AIC tends to overestimate
the correct number of latent states, especially in multivariate
cases. The classification based criteria tend to underestimate the
true number of latent states, mainly for the univariate case,
whereas their performance improves by increasing the number of
observed response variables. We also observed a significant better
behavior for NEC$_2$ with respect to the other classification based
criteria. Finally, by increasing the number of latent states the
performance of all considered criteria gets worse, mainly in the
univariate case. We conclude outlining that the results we obtained
are coherent with those observed in the literature about HM models
\citep[see, among others,][]{costa:deange:10}.

Concerning further developments of the present work, we intend to
extend the simulation study to LM models with covariates. We also
intend to rely on the most recent advances in the context of HM
models for a different formulation of the entropy that takes into
account the tendency of traditional formulation to overestimate the
uncertainty in these type of models \citep{dura:gu:12}.

\vspace{-1.5mm}
\bibliography{biblio}

\begin{thebibliography}{35}
\providecommand{\natexlab}[1]{#1}
\providecommand{\url}[1]{{#1}}
\providecommand{\urlprefix}{URL }
\expandafter\ifx\csname urlstyle\endcsname\relax
  \providecommand{\doi}[1]{DOI~\discretionary{}{}{}#1}\else
  \providecommand{\doi}{DOI~\discretionary{}{}{}\begingroup
  \urlstyle{rm}\Url}\fi
\providecommand{\eprint}[2][]{\url{#2}}

\bibitem[{Akaike(1973)}]{aka:73}
Akaike H (1973) Information theory and an extension of the {M}aximum
  {L}ikelihood principle. In: Petrov B, Csaki F (eds) Second International
  Symposium on Information Theory, Akademiai Kiado, Budapest, pp 267–--281

\bibitem[{Bartolucci(2006)}]{bart:06}
Bartolucci F (2006) Likelihood inference for a class of latent {M}arkov models
  under linear hypotheses on the transition probabilities. Journal of the Royal
  Statistical Society, series B 68:155--178

\bibitem[{Bartolucci et~al(2007)Bartolucci, Colombi, and
  Forcina}]{bart:et:al:07a}
Bartolucci F, Colombi R, Forcina A (2007) An extended class of marginal link
  functions for modelling contingency tables by equality and inequality
  constraints. Statistica Sinica 17:692--711

\bibitem[{Bartolucci et~al(2013)Bartolucci, Farcomeni, and
  Pennoni}]{bart:farc:penn:13}
Bartolucci F, Farcomeni A, Pennoni F (2013) Latent Markov Models for
  Longitudinal Data. Chapman \& Hall/CRC, Boca Raton, FL

\bibitem[{Baum et~al(1970)Baum, Petrie, Soules, and Weiss}]{baum:70}
Baum LE, Petrie T, Soules G, Weiss N (1970) A maximization technique occurring
  in the statistical analysis of probabilistic functions of {M}arkov chains.
  The Annals of Mathematical Statistics 41:164--171

\bibitem[{Biernacki and Govaert(1997)}]{bier:gova:97}
Biernacki C, Govaert G (1997) Using the classification likelihood to choose the
  number of clusters. Computing Science and Statistics 29:451--457

\bibitem[{Biernacki and Govaert(1999)}]{bier:gova:99}
Biernacki C, Govaert G (1999) Choosing models in model-based clustering and
  discriminant analysis. J Statistical Computation and Simulation 14:49--71

\bibitem[{Biernacki et~al(1998)Biernacki, Celeux, and
  Govaert}]{bier:cele:gova:98}
Biernacki C, Celeux G, Govaert G (1998) Assessing a mixture model for
  clustering with the integrated classification likelihood. PhD thesis,
  Institut National de Recherche en informatique et en automatique

\bibitem[{Biernacki et~al(1999)Biernacki, Celeux, and Govaert}]{bier:et:al:99}
Biernacki C, Celeux G, Govaert G (1999) An improvement of the {NEC} criterion
  for assessing the number of clusters in a mixture model. Pattern Recognition
  Letters 20:267--272

\bibitem[{Biernacki et~al(2000)Biernacki, Celeux, and
  Govaert}]{bier:cele:gova:00}
Biernacki C, Celeux G, Govaert G (2000) Assessing a mixture model for
  clustering with the integrated completed likelihood. {IEEE} Transactions on
  Pattern Analysis and Machine Intelligence 22:719 --725

\bibitem[{Bozdogan(1987)}]{Bozdogan1987}
Bozdogan H (1987) Model selection and {A}kaike's {I}nformation {C}riterion
  ({AIC}): The general theory and its analytical extensions. Psychometrika
  52:345--370

\bibitem[{Bozdogan(1993)}]{bozdogan1993}
Bozdogan H (1993) Choosing the number of component clusters in the
  mixture-model using a new informational complexity criterion of the
  {I}nverse-{F}isher information matrix. In: Opitz O, Lausen B, Klar R (eds)
  Information and Classification, Concepts, Methods and Applications, Springer,
  Berlin, pp 40--54

\bibitem[{Celeux and Durand(2008)}]{cele:dura:08}
Celeux G, Durand JB (2008) Selecting hidden {M}arkov model state number with
  cross-validated likelihood. Comput Stat 23:541--564

\bibitem[{Celeux and Soromenho(1996)}]{cel:sor:96}
Celeux G, Soromenho G (1996) An entropy criterion for assessing the number of
  clusters in a mixture model. Journal of Classification 13:195--212

\bibitem[{Costa and De~Angelis(2010)}]{costa:deange:10}
Costa M, De~Angelis L (2010) Model selection in hidden {Ma}rkov models: a
  simulation study. Quaderni di Dipartimento~7, Department of Statistics,
  University of Bologna

\bibitem[{Dempster et~al(1977)Dempster, Laird, and Rubin}]{demp:et:al:77}
Dempster AP, Laird NM, Rubin DB (1977) Maximum {L}ikelihood from incomplete
  data via the {EM} algorithm. Journal of the Royal Statistical Society, Series
  B 39:1--38

\bibitem[{Dias(2006)}]{dias:06}
Dias J (2006) Model selection for the binary latent class model: A {M}onte
  {C}arlo simulation. In: Batagelj V, Bock HH, Ferligoj A, Žiberna A (eds) Data
  Science and Classification, Springer Berlin Heidelberg, pp 91--99

\bibitem[{Durand and Gu\'edon(2012)}]{dura:gu:12}
Durand JB, Gu\'edon Y (2012) Localizing the latent structure canonical
  uncertainty: entropy profiles for hidden {Ma}rkov models. Tech. rep.,
  Research Report 7896, Project-Teams Mistis and Virtual Plants

\bibitem[{Fraley and Raftery(2002)}]{fraley:raft:02}
Fraley C, Raftery A (2002) Model-based clustering, discriminant analysis, and
  density estimation. Journal of the American Statistical Association
  97:611--631

\bibitem[{Hathaway(1986)}]{hattaway:86}
Hathaway RJ (1986) Another interpretation of the {EM} algorithm for mixture
  distributions. Statistics \& Probability Letters 4:53 -- 56

\bibitem[{Hernando et~al(2005)Hernando, Crespi, and
  Cybenko}]{hern:cres:cybe:05}
Hernando D, Crespi V, Cybenko G (2005) Efficient computation of the hidden
  {M}arkov model entropy for a given observation sequence. IEEE Transactions on
  Information Theory 51:2681--2685

\bibitem[{Kass and Raftery(1995)}]{kass:raftery:1995}
Kass RE, Raftery AE (1995) Bayes factors and model uncertainty. Journal of the
  American Statistical Association 90:773--795

\bibitem[{Langeheine(1994)}]{lang:94}
Langeheine R (1994) Latent variables {M}arkov models. In: von Eye A, Clogg C
  (eds) Latent variables analysis: Applications for developmental research,
  Sage, Thousand Oaks, CA, pp 373--395

\bibitem[{Langeheine and Van~de Pol(1994)}]{lang:vand:94}
Langeheine R, Van~de Pol F (1994) Discrete time mixed {M}arkov latent class
  models. In: Dale A, RB D (eds) Analyzing social and political change. A
  casebook of methods., London: Sage., pp 171--197

\bibitem[{Li(2005)}]{li:05}
Li J (2005) Clustering based on a multilayer mixture model. Journal of
  Computational and Graphical Statistics 14:547--568

\bibitem[{Lin and Dayton(1997)}]{lin:day:97}
Lin TH, Dayton CM (1997) Model selection information criteria for non-nested
  latent class models. Journal of Educational and Behavioral Statistics
  22:249--264

\bibitem[{Magidson and Vermunt(2001)}]{magi:verm:01}
Magidson J, Vermunt JK (2001) Latent class factor and cluster models, bi-plots
  and related graphical displays. Sociological Methodology 31:223--264

\bibitem[{McLachlan and Peel(2000)}]{mcla:peel:00}
McLachlan G, Peel D (2000) Finite Mixture Models. Wiley

\bibitem[{Nylund et~al(2006)Nylund, Asparouhov, and Muth\'{e}n}]{nyl:et:al:07}
Nylund K, Asparouhov T, Muth\'{e}n B (2006) Deciding on the number of classes
  in latent class analysis and growth mixture modeling: A monte carlo
  simulation study. Structural Equation Modeling 14:535--569

\bibitem[{Schwarz(1978)}]{schw:78}
Schwarz G (1978) Estimating the dimension of a model. The Annals of Statistics
  6:461--464

\bibitem[{Tierney and Kadane(1986)}]{tiern:kada:86}
Tierney L, Kadane JB (1986) {Accurate Approximations for Posterior Moments and
  Marginal Densities}. Journal of the American Statistical Association
  81:82--86

\bibitem[{Welch(2003)}]{welch:2003}
Welch LR (2003) Hidden {M}arkov models and the {B}aum-{W}elch algorithm. IEEE
  Information Theory Society Newsletter 53:1--13

\bibitem[{Wiggins(1973)}]{Wig:73}
Wiggins L (1973) Panel analysis. Latent probability models for attitude and
  behavior processes. Elsevier, New York, US

\bibitem[{Wolfe(1970)}]{wolfe:1970}
Wolfe JH (1970) Pattern clustering by multivariate mixture analysis.
  Multivariate Behavioral Research 5:329--350

\bibitem[{Zucchini and MacDonald(2009)}]{zucc:macd:09}
Zucchini W, MacDonald IL (2009) Hidden {M}arkov Models for time series: an
  introduction using R. Springer-Verlag, New York

\end{thebibliography}

\bibliographystyle{spbasic}
\end{document}